\documentclass[twocolumn,showpacs,preprintnumbers,amsmath,amssymb]{revtex4}
\usepackage{graphicx}
\usepackage{dcolumn}
\usepackage{bm}

\begin{document}
\preprint{APS}
\title{Light-induced atomic desorption and diffusion of Rb from porous
alumina}
\author{S. Villalba, H. Failache and A. Lezama}
\affiliation{Instituto de F\'{\i}sica, Facultad de Ingenier\'{\i}a,
Universidad de la Rep\'{u}blica,\\ J. Herrera y Reissig 565, 11300
Montevideo, Uruguay}
\date{\today}

\begin{abstract}
We present the first study of light induced atom
desorption (LIAD) of an alkali atom (Rb) in porous alumina.  We observe the variation due to LIAD of the rubidium density in a vapor cell as a function of illumination time, intensity and wavelength. The simple and regular structure of the alumina pores allows a description of the atomic diffusion in the porous medium in which the diffusion constant only depends on the known pore geometry and the atomic sticking time to the pore wall. A simple one-dimensional theoretical model is presented which reproduces the essential features of the observed signals. Fitting of the model to the experimental data gives access to the diffusion constant and consequently the atom-wall sticking time and its dependence on light intensity and wavelength. The non-monotonic dependence of the LIAD yield on the illumination light frequency is indicative of the existence of Rb clusters in the porous medium.
\end{abstract} 

\maketitle
\section{\label{introduccion}Introduction}
In a spectroscopic cell containing an alkali atom vapor, a
substantial fraction of the atoms are adsorbed on the cell walls. At
steady state, the gas density is in equilibrium with the adsorbed
atomic fraction. In some cells, depending on the cell material or coating, when the cell is illuminated with moderate intensity (1 - 1000 mW/cm$^{2}$)
nonresonant light, a significant increase in the atomic vapor
density is produced as a consequence of the release of atoms from
the cell surface into the gas phase. Such effect has been named
light induced atomic desorption (LIAD) \cite{Gozzini1993}.\\
LIAD has received considerable attention in recent years due to its
application as a light-controlled atom dispenser under high vacuum
conditions. Such dispenser has been successfully used to load
magneto-optical atom traps \cite{Anderson2001,Atutov2003,Klempt2006} and hollow optical fibers \cite{Ghosh2006,Londero2009,bhagwat2009}. Its use has also been considered for atomic magnetometers, gyroscopes and clocks \cite{Karaulanov2009, Bogi2009}. In
addition, LIAD has attracted the attention of astrophysicists since
it has been related to the observed abundance of alkaline elements
in nonpermanent
extraterrestrial atmospheres \cite{Yakishinskiy2003}.\\
LIAD is understood as a non-thermal effect as opposed to light
desorption produced with high power sources in which a significant
heating results from light absorption by the substrate. In
poly-dimetilsiloxane (PDMS) \cite{Xu1996}, paraffin
\cite{Alexandrov2002} and sapphire \cite{Bonch1990} a frequency
threshold in the infrared, similar to that of the photoelectric
effect on metals, has been observed. Also, an increasing efficiency
of LIAD with light frequency has been reported in several
samples \cite{Alexandrov2002,Atutov1999,Mariotti1996}.\\
LIAD was first observed in sodium vapor glass cells in which the
inner cell walls were coated with a thin layer of PDMS. The effect
was also observed with K \cite{SGozzini2004}, Rb and Cs atoms
\cite{Alexandrov2002} (sometimes in the presence of a buffer gas).
Initially, it was considered that LIAD was specific to PDMS coatings
\cite{Xu1996}. However, LIAD was later-on reported in cells coated
with different polymers such as octadimethyl-cyclotetrasiloxane
(OCT) \cite{Xu} and paraffin \cite{Alexandrov2002}. LIAD has also
been observed on several uncoated surfaces such as glass \cite{Klempt2006,bhagwat2009}, stainless
steel \cite{Klempt2006} and sapphire \cite{Bonch1987}. More
recently, LIAD has been studied
in porous amorphous materials such as porous silica \cite{Burchianti2004}.\\
All observations of LIAD in porous or coated surfaces present some common features such as the
characteristic time scale of the atomic desorption (several
seconds). However, other aspects may vary significantly between
different atomic species and coatings depending also in the cell
preparation procedure. In particular, large variations are observed
in the desorption yield. In cells coated with PDMS, LIAD may result
in an increase of the atomic gas phase density of  several orders of
magnitude \cite{Xu1996} while only density-increase-factors of a few
units were reported for paraffin \cite{Alexandrov2002}. Smaller factors were
observed on uncoated surfaces as in the present study. The question
on whether there is a common mechanism underlying all LIAD
observations is still open \cite{Hamers2005}.\\
The first tentative explanation of LIAD at the microscopic level
was suggested by Xu et al \cite{Xu1996}. The mechanism involves the
modification by light of the weakly bonded chemical complex
formed between a PDMS molecule and the Na atom or Na$_{2}$ molecule.
More recently, this mechanism was further investigated through the
measurement of the thermal distribution of desorbed atom velocities
\cite{Brewer2004}. This interpretation of  LIAD is consistent with the
observation of a threshold light frequency for LIAD in PDMS but leaves
unexplained several aspects of its dynamics. As discussed by Atutov
et al \cite{Atutov1999}, in addition to the atomic desorption from the
surface, the diffusion of the atoms within the surface coating plays
an essential role in the temporal evolution of LIAD. To a large
extent, LIAD in coated surfaces is a consequence of light-induced
modification of  the atomic mobility and diffusion within the
coating polymer. Atutov et al  \cite{Atutov1999} have modelled such process
assuming a phenomenological dependence of the diffusion coefficient
on light.  Alexandrov et al have described the LIAD
dynamics with the help of rate equations with a light dependent term
representing the flux of atoms from the coating into the gas phase \cite{Alexandrov2002}. Recently, the model of LIAD in coated surfaces suggested by Atutov has been revised and improved by Rebilas and co-workers \cite{Rebilas2009_1,Rebilas2009_2}.\\
LIAD from uncoated dielectric surfaces, such as sapphire or glass,
deserves special consideration. In such systems, alkaline atoms can
be individually adsorbed on the surface or agglomerated into
clusters. The presence of clusters may result in a visible change of
the sample transparency or even in coloration \cite{Burchianti2008}. Blue-green coloration by Rb
of otherwise transparent (or white) samples  has been observed
in several experiments including the ones described here. The role
of the light in these samples is double since it can produce the
direct desorption of the atoms from the dielectric surface and the evaporation of the
atomic clusters \cite{Burchianti2006,Burchianti2009}. Also, under suitable conditions, the light
may also control the growth of the clusters from atoms in the vapor
phase. Such conditions are favored in porous media where the
desorbed atoms remain confined and available to participate in the
cluster regrowth. A characteristic feature of the LIAD involving
cluster evaporation, is the non-monotonic dependence of the
desorption yield on the light frequency. Such behavior is
interpreted as the consequence of resonant surface plasmon excitation in the cluster \cite{Burchianti2006}. A second characteristic of
these systems is its ``memory". The response  strongly
depends on the illumination history including the timing of the
bright and dark periods and the corresponding color sequence
\cite{Burchianti2006}.\\
A common feature governing the dynamics of both, the LIAD in polymer
coated surfaces and in porous dielectrics, is the successive
occurrence of two distinctive processes: i) atomic desorption from
the surface (or cluster) ii) diffusion in the intermediate medium
(either the polymer or the porous medium) prior to the atom release
in the vapor phase. A precise modelling of LIAD should involve the
simultaneous account of these two processes. The desorption
mechanism is at present only qualitatively understood \cite{Xu1996,Brewer2004,Burchianti2006}.
Also, little understanding is currently available on the mechanisms
determining the variation with light of the atomic mobility in the
polymer coating. Diffusion in porous silica is presumably simpler
since the atomic motion inside the pores may be assumed to occur in
a diluted vapor. However, the random nature of the pore geometry
complicates the modelling of such process.\\
In this paper we present the experimental study and theoretical
modelling of LIAD with Rb atoms adsorbed in thin membranes of porous
alumina. The porous medium is produced by anodization of aluminum
and results in a very regular array of cylindrical pores with small
size dispersion. The average diameter of the pore tube is 200 nm and
its length 60 $\mu$m. In consequence, the pore geometry is well
known and this allows a simple and accurate description of the
atomic diffusive motion in the pore. The
diffusion process is the result of a random sequence of atomic free flight of the atoms confined within the pore wall.  After a collision with the wall, the atom sticks to
the wall for some time after which it is desorbed again flying with
random direction and velocity. Under such assumptions, the diffusive
part (ii) of the LIAD dynamics, including  the release of atoms into
the gas phase, is determined by the pore geometry and the atom-surface sticking time. Such
picture allows us to model the atomic diffusion in the pores as well
as the atom exchange between the porous material and the outside gas
volume.  The resulting LIAD dynamics is mainly dependent on the
atomic diffusion constant in the porous medium which can be obtained
from the fitting of the observed variation of the atomic density in
the gas phase. From the knowledge of the diffusion constant, the atom-wall sticking time can be determined. We have
investigated the variation of the sticking time with light
intensity and color.\\
\section{\label{Montaje}Experiment}
\begin{figure}
\includegraphics[width=8.6cm]{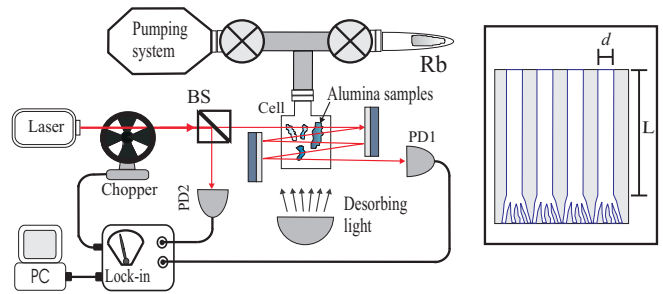}
\caption{\label{fig:montaje_alumina} (Color online) Experimental setup. Inset: Schematic cut of the alumina membrane for a plane parallel to the pores. BS: beam splitter, PD1, PD2: photodetectors.}
\end{figure}

We have used porous alumina membranes manufactured by
Whatman International Limited. The circular flat membranes have a
diameter of one inch and a thickness of 60 $\mu$m. The membrane is
traversed by a regular array of cylindrical hollow tubes with 200 nm
diameter. The tubes form a honey comb like array with a pore density of $10^{9}/$cm$^{2}$. The diameter of the pores
are uniform over most of their length. On one side of the membrane,
along 1 $\mu$m, the pores divide into several smaller branches with
20 nm typical diameter (see the inset in Fig.
\ref{fig:montaje_alumina}). Before contact with the Rb vapor the
porous membranes are translucid and white. In order to fit into the
vacuum glass cell, the membranes are divided in pieces of typically
0.5 cm$^{2}$.\\
The experimental setup is shown in Fig. \ref{fig:montaje_alumina}.
We have used a vacuum glass cell with 2.5 x 2.5 x 4.5 cm dimensions.
The cell is connected via a glass-to-metal transition fitting to an
ion pump and a metallic Rb reservoir. The Rb density in the glass
cell is monitored by measuring the absorption of a laser beam issued
from an extended cavity diode laser. Using a saturated absorption
setup, the laser frequency is stabilized to the  $^{85}$Rb $\mathrm{F}=3\mapsto\mathrm{F'}=4$
transition in the D2 line (780 nm). In order to increase the absorption signal, the laser
beam crosses the cell several times.  We have used a balanced
detection scheme to reduce sensitivity to laser intensity
fluctuations. Half of the laser power is sampled before the cell and
detected with a photodiode. A second photodiode monitors the
intensity of the beam transmitted through the cell. The outputs of
the two photodiodes are subtracted. In order to eliminate noise from
ambient light, including the light used for the LIAD, the laser beam
is modulated with a chopper and lock-in detected. The
illumination of the porous alumina samples is made with high power
LEDs ($100$ mW) in order to have a non-thermal source with a well
defined spectrum. Three different LEDs were used centered at
$455$, $504$ and $617$ nm (typical spectral width 10 nm). An optical
arrangement (not shown in Fig. \ref{fig:montaje_alumina}) allows a
uniform
illumination of the porous sample by the LED light.

Prior to the introduction of the porous alumina membrane, the glass
cell was evacuated (10$^{-6}$ torr) and baked for several hours at 300 C. Such
precaution appeared to be essential since we have observed
significant LIAD from the unbaked cell presumably due to some
uncontrolled coating. After the baking procedure, the LIAD from the
cell walls was negligible. Following the cell cleanup, several
pieces of the alumina membrane were introduced and vacuum baked for
several days at 150 Celsius. The pieces of alumina lied on the cell
bottom. We had no control on the side of the membrane that faces the
cell wall, so some of the pieces present the largest pore apertures toward
the cell bulk volume while others present the narrow ramification ends. After the initial cleanup of the alumina, the cell was
returned to room temperature and the valve separating the cell from
the metallic Rb reservoir opened. Keeping the Rb reservoir and the
vacuum connecting tubes slightly heated ($\sim 50$ C), the Rb was allowed to
diffuse into the cell and the porous alumina. After a few days, a
visible blue coloration appeared in the alumina indicating the
presence of Rb. After a sufficiently long period all the samples
were dark blue. However two different blue tones were observed among
the samples. We interpret such difference as a consequence of the
two possible orientations of the membrane pieces with respect to the
cell wall. The blue coloration is an indication of the formation of
Rb clusters \cite{Burchianti2008}. We have checked that the cluster formation is
entirely reversible. The original white coloration of the alumina
could be recovered after pumping the cell during a few hours while
illuminating with an incandescent lamp.\\
\section{\label{Montaje}Experimental results}
\begin{figure}
\includegraphics[width=9.0cm]{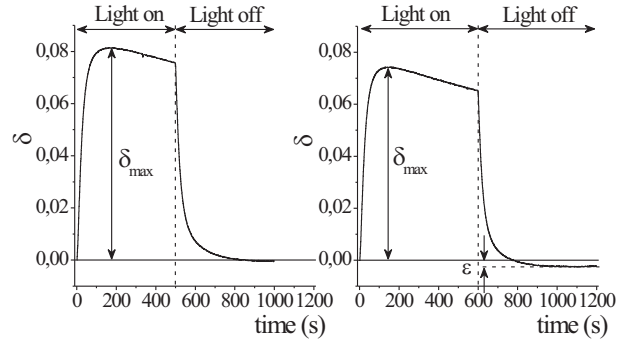}
 \caption{\label{Grafcarac} Typical observed variations of the
 relative atomic vapor density for two different illumination-time intervals.}
\end{figure}

We have observed the LIAD of Rb from the porous alumina by
monitoring the laser absorption in the cell bulk while turning on
and off the illumination by a LED. We have recorded the relative
variation of the vapor density
$\delta(t)\equiv(\rho(t)-\rho_{0})/\rho_{0}$ as a function of time
where $\rho(t)$ is the density of Rb in the cell and
$\rho_{0}$ the equilibrium density in the dark. Fig. \ref{Grafcarac}
shows two typical records obtained with the same illumination for
two different light-on intervals (500 and 600 s). In general, the
relative density reaches a maximum $\delta_{max}$ after a few tens
of seconds depending on light intensity. After that the Rb density
slowly decreases towards a new steady state in the presence of
light. When the light is turned off, $\delta$ decreases on a time
scale comparable to the rise time. Two different
behaviors have been observed for long times after the light
switching off. Either the density returns monotonically to the
initial equilibrium density $\rho_{0}$ or drops below $\rho_{0}$ by
an amount $\varepsilon$ (see Fig. \ref{Grafcarac}) after what it slowly
grows towards $\rho_{0}$. The later behavior is observed if the
light
intensity and the illumination interval are sufficiently large.\\

\begin{figure}
\centering
\includegraphics[width=8.6cm]{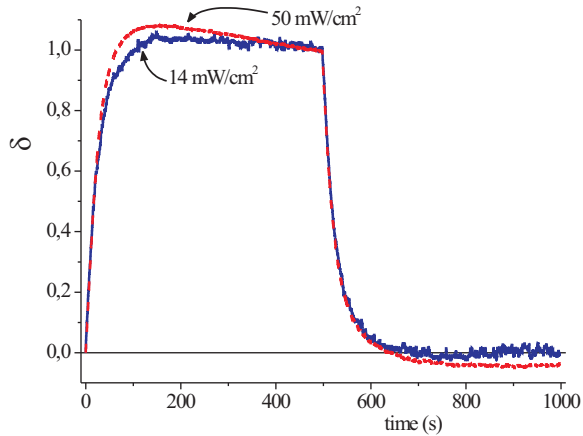}
\caption{\label{Ampli-Forma_Int} (Color online) Two records of the
relative atomic density variation illustrating the difference in shape for
different illumination intensities. The traces have been re-scaled to signal the difference in shape.}
\end{figure}
Fig. \ref{Ampli-Forma_Int} shows the evolution of the Rb density
in the cell for two different illumination intensities. Notice the
variation in the shape of the trace. Similar shape variations were
also observed in experiments with porous silica
\cite{Burchianti2004}, although not
reproduced by the proposed theoretical models.\\

\begin{figure}
\includegraphics[width=8.6cm]{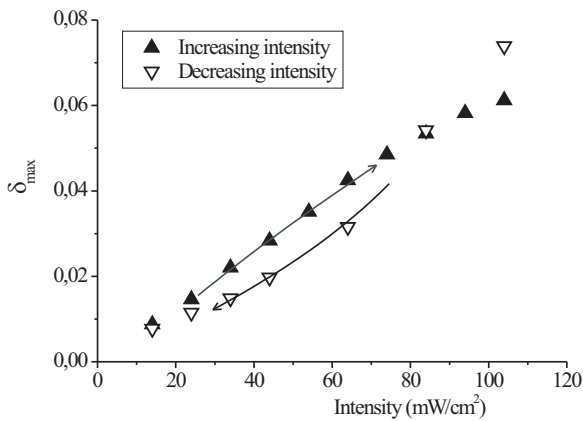}
\caption{\label{fig:may9amplitazul} Maximum relative atomic density
$\delta_{max}$ as a function of the illumination intensity. The
measurements were registered with a sequence of illumination
intervals of $140 s$ followed by intervals of $600 s$ without
illumination. Solid(hollow) triangles: increasing(decreasing) illumination intensity.}
\end{figure}

We have observed that the efficiency of the LIAD process depends on
the porous alumina history, as was also noticed in other systems
\cite{Alexandrov2002,Burchianti2006}. A monotonic reduction in the
maximum relative Rb density variation $\delta_{max}$ is observed for
several successive illumination cycles keeping constant the light
intensity. In addition, as the intensity is changed between successive illumination periods, the signal variation is different depending on whether the light intensity is increased or decreased (see Fig. \ref{fig:may9amplitazul}). For low enough light intensities the system is not
 appreciably modified by the
illumination and a linear dependence of the LIAD yield on light intensity is observed. The nonlinear dependence, visible in
Fig.\ref{fig:may9amplitazul} for large intensities, can be
attributed to the depletion of the available Rb inside the nano-pores.\\

\begin{figure}
\includegraphics[width=7.6cm]{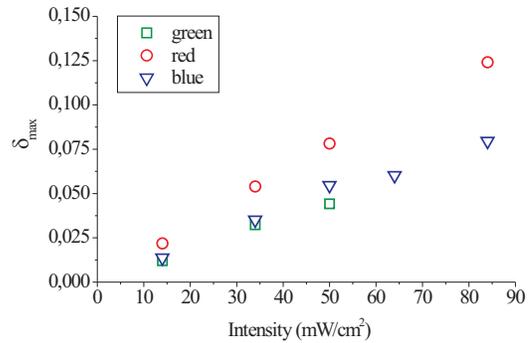}
\caption{\label{fig:GampconI} (Color online) Maximum relative atomic density
$\delta_{max}$ as a function of the illumination intensity for
different illumination wavelengths ($455$, $505$ and $617$ nm).}
\end{figure}

Fig. \ref{fig:GampconI} shows $\delta_{max}$ as a function of
illumination intensity for three different wavelengths. The
measurements were taken alternating the three available light colors
successively for each intensity. The effect of the history on the
LIAD efficiency is so reduced for the comparison among measurements
taken with different wavelengths. In Fig.\ref{fig:GampconI} the
non-linear variation of $\delta_{max}$ is only noticeable for the
highest
intensities.\\
From the linear fit of the data in Fig. \ref{fig:GampconI} one can
evaluate, for each wavelength, the coefficient
$\alpha_{\lambda}\equiv \delta _{max} \hbar \omega /I$ proportional
to the LIAD desorption rate per photon flux.  We observe that this coefficient for blue, green and red light varies in proportion to 1, 0.73 and 1.1 respectively. Such result indicate a
non-monotonic evolution of the LIAD yield with photon
energy. \\
\section{\label{model}Theoretical model}

We model the evolution of the atomic density inside the cylindrical
pores as a one-dimensional diffusion process. The typical sticking time $\tau_s$ of alkali atoms on dielectric surfaces is of the order of tens to
hundreds microseconds. After desorption, the atoms leave the
internal surface of the pore with thermal velocity in a random
direction with a Lambertian probability distribution
\cite{Goodman1976}. The gas density inside the pores is considered sufficiently low to neglect  the collisions between flying atoms. At room temperature and for hundred nanometers
tube diameters, after a few nanoseconds flight, the atom is
again adsorbed on the pore surface. Since the pore length is much
larger than its diameter, we can consider that the atoms execute a
one dimensional random walk, along the pore axis, characterized by
the diffusion constant (see Appendix):
\begin{equation}\label{ec:dif-geometria}
D=\frac{l^{2}}{2\tau} =\frac{d^{2}}{3\tau}
\end{equation}
Where $l^{2}$ is the mean square displacement per step in the random
walk, $\tau$ is the mean interval between steps which is essentially
determined by the sticking time $\tau\simeq\tau_s$ on the internal
pore surface and $d$ is the pore diameter.

The atomic desorption is described by a reduction of $\tau_{s}$
induced by the light. We assume a simple linear dependence:
\begin{equation}\label{ec:tiempostick-luz}
\tau_{s}=\tau_{s0}(1-\kappa I)
\end{equation}
where $\tau_{s0}$ is the sticking time in the dark, $I$ is the light
intensity and $\kappa$ a coefficient which is wavelength
dependent. In consequence:
\begin{equation}\label{}
D=\frac{D_{0}}{(1-\kappa I)}
\end{equation}
with $D_{0}=d^{2}/(3\tau_{s0})$ being the atomic diffusion
constant in the dark.\\

\begin{figure}
\includegraphics[width=7.6cm]{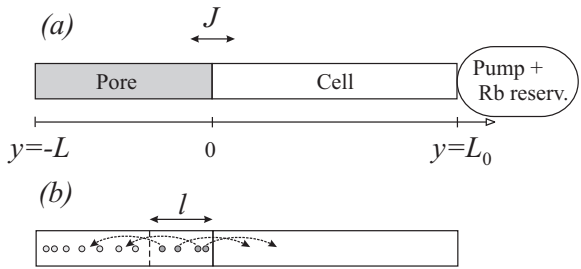}
\caption{a) Schematic one-dimensional model for the system. b)
Illustration of the atomic release form the pore end into the gas
cell ($J^{+}$ in Eq. \ref{J+}).}\label{fig:modelo1dim}
\end{figure}

Fig. \ref{fig:modelo1dim} present a scheme of the one-dimensional
model of the system. The cylindrical pore, considered closed on
its left end has a total length $L$. To the right of the pore, the
atomic vapor cell, associated to a length $L_{0}$, is
connected to a reservoir accounting for the vacuum pumping and the
external Rb reservoir.\\

The (linear) density of atoms $\mu(y,t)$ inside the pore is described by
the diffusion equation
\begin{equation}\label{}
\frac{\partial \mu}{\partial t}=D\frac{\partial^2 \mu}{\partial y^{2}}
\end{equation}
where $y$ is the position coordinate inside the pore (see
Fig.\ref{fig:modelo1dim}).

The total number of atoms $N$ in the cell is $N=N_{g}+N_{w}$ where
$N_{g}$ represents the atoms in the gas phase and $N_{w}$ the atoms
adsorbed to the cell wall. The fraction of atoms in the gas phase
relative to the total number of atoms is assumed to be a constant
for given temperature and illumination conditions \cite{Stephens1994}:
\begin{equation}\label{MyMg}
\frac{N_{g}}{N}=\frac{L_{0}}{L_{0}+\Delta}
\end{equation}
Here $\Delta$ represents an effective cell length corresponding to
adsorbed atoms. Since the sticking time of the atoms to the cell
walls can depend on light intensity, we consider that $\Delta$
depends on the illumination in the form:
$\Delta=\Delta_0(1-\zeta I)$ where $\zeta$ is a
coefficient that can depend on wavelength. The evolution of the atom number
$N$ in the cell is described by the equation
\begin{equation}\label{dM}
\frac{dN}{dt}=\frac{dN_g}{dt} \left\{
1+\frac{\Delta}{L_0}\right\}=J-\gamma(N_g-N_{g0})
\end{equation}
where $J$ is the net atomic flux at the pore-vapor interface. The
rate $\gamma$ describes the return to the equilibrium atom number
$N_{g0}$ determined by the external pumping system and Rb reservoir.

We separate the net flux $J$ into two contributions  $J=J^++J^-$
describing the atoms leaving and entering the pore respectively. The flux of atoms entering the pores from the cell gas is given by:
\begin{equation} \label{J-}
J^{-} =-\frac{\overline{v}}{2L_0}N_{g},
\end{equation}
where $\overline{v}\equiv \langle |v_{y}|\rangle$ is the mean
magnitude of the atomic velocity in the direction of the pore.
The simple geometry of our system allows the evaluation of $J^+$ without additional assumptions by considering that the atoms within a mean step
length $l$ from the pore end have a probability $1/2$ for leaving
the pore in the time interval $\tau$ (see Fig. \ref{fig:modelo1dim}
b), then
\begin{equation} \label{J+}
J^{+}=[\mu(0)l]\frac{1}{2}\frac{1}{\tau}\simeq\mu(0)\frac{D}{l}.
\end{equation}
The equations describing the evolution of the atomic densities $\mu$
and  $n\equiv N_g/L_0$  inside the pores and in the cell gas phase
respectively are:
\begin{subequations}\label{ecuaciones}
\begin{eqnarray}
 \frac{\partial \mu}{\partial t} &=& D
  \frac{\partial^2 \mu}{\partial y^{2}} \\
 \nonumber
  \frac{d n}{d t} &=& \frac{D}{l L_{c}(1-\sigma I)}\mu(0)-\frac{\left(\tilde{\gamma}+\frac{\overline{v}}
  {L_{c}}\right)}{(1-\sigma I)}n \\
  & & +\frac{\tilde{\gamma}}{(1-\sigma I)} n_{0}
\end{eqnarray}
\end{subequations}
where we have introduced the parameters $\sigma
\equiv\frac{\zeta}{\left(1+\frac{L_{0}}{\Delta_{0}}\right)}$,
$L_c\equiv (L_{0}+\Delta)$ and
$\tilde{\gamma}\equiv\frac{\gamma_{0}L_{0}}{Lc}$. $n_0$ is the equilibrium value of the atomic
density in the vapor cell.

The steady state densities in the dark inside the pores $\mu_0$ and in the vapor cell $n_0$ are linked through the condition:
\begin{equation}\label{estacionario}
J=\frac{D_{0}}{l}\mu_0-\frac{\overline{v}}{2} \: n_0=0.
\end{equation}

The boundary conditions at the pore ends are (see Eqs. \ref{J-} and \ref{J+}):

\begin{subequations}\label{borde}
\begin{eqnarray}
  -D\frac{\partial \mu}{\partial y}
  \biggr| _{y=0} &=& \frac{D}{l}\mu(0)-\frac{\overline{v}}{2}\;n\\
  \frac{\partial \mu}{\partial y}
  \biggr| _{y=-L} &=& 0
\end{eqnarray}
\end{subequations}


From the above equations, it is possible to derive an approximate relation between the observed variation of the gas density in the cell and the corresponding change in the diffusion constant inside the pores. For this we notice that in our
system, the return to equilibrium (under constant illumination)
occurs on time scale which is long compared to the observation time. One
can then consider that during the LIAD the total atomic population
(inside the pores and in the cell) remains approximately constant:
\begin{equation}\label{sinLuz}
\mu L + n L_c \simeq(\frac{l\;\overline{v}}{2D_{0}} L+L_c)n_{0}
\end{equation}
where we used Eq. \ref{estacionario}.

When the sample is illuminated, the LIAD effect redistribute the
atoms along the pore in a characteristic diffusion time $L^2/2D$.
If we assume that the gas phase density reaches its maximum $n_{max}$ in a
time which is long compared to the diffusion time, one can consider
that the corresponding atomic density inside the pores is
approximately uniform $\mu\simeq\mu_{max}$. Using Eqs. \ref{dM} (with $\gamma =0$)
and \ref{sinLuz} we have:

\begin{equation}\label{dm}
\delta_{max}\approx\frac{n_{max}-n_{0}}{n_{0}}=\frac{\Delta
D}{D_{0}}\frac{1}{\left(1+\frac{2DL_{c}}{\overline{v}\;l\;L}\right)}
\end{equation}

Eq. \ref{dm} can be used for a quick estimate of the relative variation of the diffusion constant from the observed change in the vapor density, provided the second term inside the brackets in Eq. \ref{dm} is small. In the conditions of our experiment such term is of the order of unity.\\

Some of the parameters appearing in the model can be directly
determined for our system. From the porous alumina manufacturer we
know that $L=60\mu m$, and $d=200nm$. In consequence, $\langle l^{2} \rangle=\frac{2}{3}d^{2}\approx1.6\times10^{-7}
m^2$. The mean atomic velocity at room temperature is
$\overline{v}\approx 140m/s$. The other parameters are determined
through least square fitting of the numerical model to the
experimental data. For this, we have numerically integrated the differential equations \ref{ecuaciones} with the boundary conditions given in Eqs. \ref{borde}.\\

\begin{figure}[h]
\centering
\includegraphics[width=8.6cm]{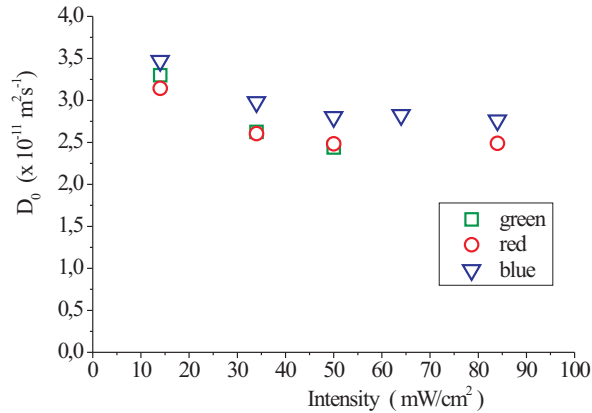}
\caption{(Color online) Values of the diffusion constant D$_{0}$ obtained from the
fitting of different experimental traces obtained with three
illumination colors.}\label{dispersion_color}

\end{figure}

Fig. \ref{fig:ParamAjust_set7Fin} shows a typical
experimental register together with the corresponding signal
calculated from the model. The values of the parameters obtained
from the fitting are presented in Table. \ref{tab:paramajuste}. The
given uncertainties correspond to the scattering of the results of the
fitting for different experimental runs. The value of $D_{0}$ given
in Table \ref{tab:paramajuste} results from the average of the data
obtained with all three excitation wavelength. Interestingly enough,
the plot of the fitted values of $D_{0}$ for different runs reveal a
systematic grouping for each of the three colors used for LIAD (see
Fig. \ref{dispersion_color}). In our model, $D_{0}$ corresponding
to the diffusion constant in the dark, is taken as constant and
independent of the desorbing light color. However, the grouping
observed in Fig. \ref{dispersion_color} may reveal a dependence of
$D_{0}$ on the illumination history. Such feature could be an
indication of cluster formation and cluster-light interaction. The investigation of cluster formation is outside the scope of this work.

We notice that the value of $L_{0}$ in Table \ref{tab:paramajuste} is large compared to the length  ($\lesssim 10m$) estimated from the actual glass cell volume. However, the total effective volume available to the atoms outside the porous alumina also depends on the vacuum system tubes and surfaces  \cite{Karaulanov2009}. The parameter $\sigma$ reflects the dependence of the effective cell length on the illumination. The numerical fitting is quite insensitive to this parameter giving a large scattering of the results. The uncertainty in this parameter prevents the determination of a wavelength dependence. On the other hand, different values of the coefficient $\kappa$ are obtained depending on the wavelength of  the illuminating light.

From the parameters in Table \ref{tab:paramajuste} we can check that the assumptions made for the derivation of Eq. \ref{dm} are reasonable for our system. The estimate of the maximum relative vapor density variation obtained using Eq. \ref{dm} only differs in a few percents from the value resulting from the numerical integration of Eqs. \ref{ecuaciones}.

\begin{figure}[h]
\centering
\includegraphics[width=7.6cm]{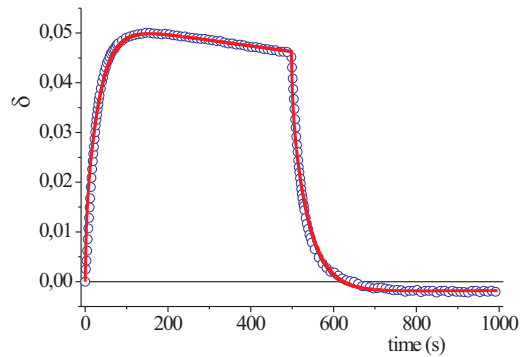}
\label{fig:ParamAjust_set7Fin} \caption{(Color online) Example of the fitting of the calculated signal (solid line) to the experimental data
(circles).}\label{fig:ParamAjust_set7Fin}
\end{figure}

\begin{table*}
\caption{\label{tab:paramajuste} Fitted values of the parameters of the model.}
\begin{ruledtabular}
\begin{tabular}{cccccccc}
& D$_{0}[m^{2}.s^{-1}]$& L$_{0}[m]$ & $\gamma[s^{-1}]$ & $\sigma[mW^{-1}]$ & $\kappa_{red}[mW^{-1}]$ & $\kappa_{green}[mW^{-1}]$ & $\kappa_{blue}[mW^{-1}]$ \\
\\
& $2.8 \pm 0.5 \times10^{-11}$ & $106\pm 33$ & $2.4 \pm 1.5 \times10^{-4}$ & $2.2\pm 2.0 \times10^{-3}$ & $7.8 \pm 0.7 \times10^{-3}$& $4.6 \pm 0.5 \times10^{-3}$ & $5.2 \pm 0.5 \times10^{-3}$
\\
\end{tabular}
\end{ruledtabular}
\end{table*}

\begin{figure}[h]
\centering
\includegraphics[width=8.6cm]{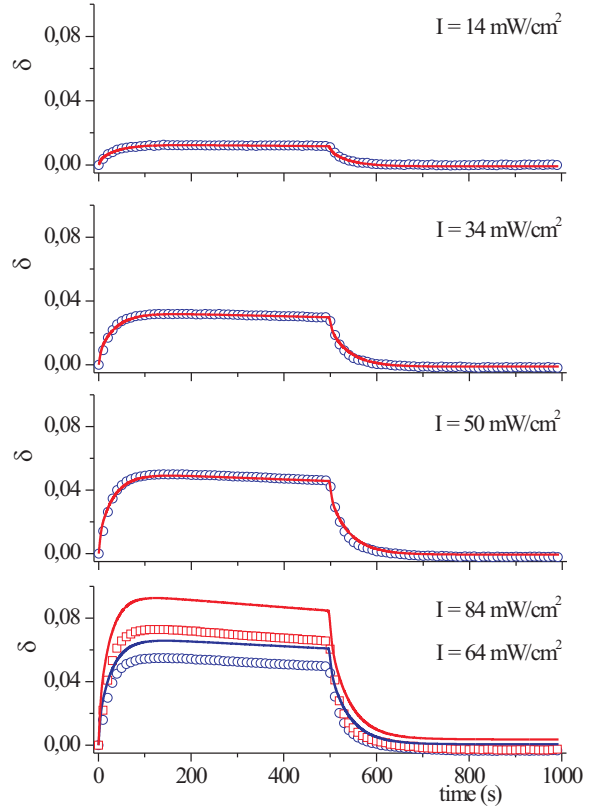}
\caption{(Color online) Comparison of the experimental data (circles and squares) with the
calculated signal (solid lines) obtained with the parameters in Table
\ref{tab:paramajuste} for $\lambda=455$nm }\label{AjusteModeloAzul}
\end{figure}

A comparison of the predictions of the theoretical model with the
experimental observations is given in Fig. \ref{AjusteModeloAzul}
for traces obtained with blue desorbing light. Except for the
largest intensities, where the effects of saturation and illumination history are
expected to be significant, the model correctly describes the growth
of the LIAD signal with light intensity. Similar results are
obtained for the other colors used for illumination.

\begin{figure}[h]
\centering
\includegraphics[width=8.6cm]{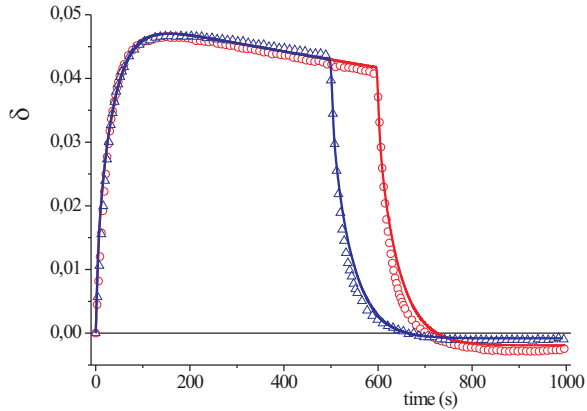}
\caption{(Color online) Observed (circles and triangles) and calculated (solid line) signals for two illumination intervals.}\label{EpsilonconTiempLuz}
\end{figure}

\begin{figure}[h]
\centering
\includegraphics[width=8.6cm]{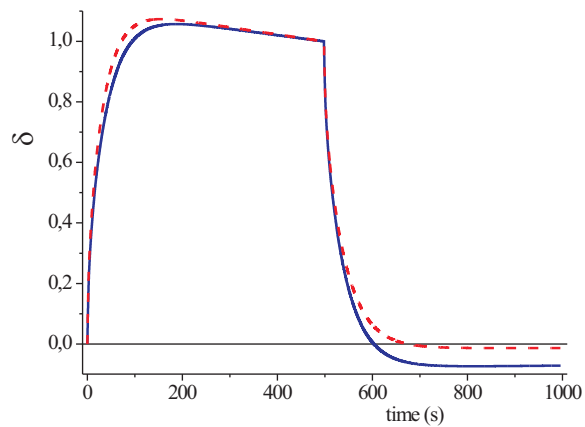}
\caption{(Color online) Calculated traces for different illumination intensities.
The plots have been re-scaled to signal the difference in shape.}
\label{calculoNormaliz}
\end{figure}

Our theoretical model appears to correctly account for several
features of the experimental signal. As shown in Fig.
\ref{EpsilonconTiempLuz} the signal shape variation as a function of
the illumination time interval is well described. In particular, the
``undershoot'' $\varepsilon$ of the vapor density below the initial
density is well reproduced. Such ``undershoot'' is due to the small
variation of the total number of atoms (due to the external pumping
system) during illumination. As the illumination is turned off, the
atoms are rapidly re-adsorbed by the porous alumina in a time
shorter than the one required to equilibrate the cell with the
external pump and Rb supply. The model also reproduces the
difference in shape of the temporal evolution between low and large
illumination intensities as shown in Fig \ref{calculoNormaliz}. Such
shape variations were previously observed but not reproduced by
existing models \cite{Burchianti2004}.

From the measured value of $D_{0}$ using Eq. \ref{ec:dif-geometria}
one can determine the sticking time $\tau_{s}$ of the atoms to the
pore walls. The obtained value $\tau_{s0}\simeq 500 \mu s$
lies within the range of previous observations for alkali atoms on
dielectric surfaces. A summary of the
sticking times reported in the literature for several alkali atoms
and surfaces is presented in Table \ref{tab:table}. The value of
$\tau_{s}$ is several orders of magnitude larger than the mean
time-of-flight  of the atoms between collisions with the pore walls
$\tau_{0}\sim 1ns$. At a given time, the fraction of atoms in
the gas phase inside the pores relative to the total number of atoms participating in the diffusion is $\tau_{0} / \tau_{s}\sim 10^{-5}$. From the
values in Table \ref{tab:paramajuste} we estimate a relative
variation of the atomic gas density within the pores of 60\% for
illumination with 50 mW/cm$^{2}$ of red light.

\section{\label{conclusiones}Conclusions}

We have studied LIAD of Rb atoms contained within alumina nanopores.
We observed, as a function of time, the variations of the Rb
density in the cell surrounding the porous alumina as illuminating
light with different colors is turned on and off. We have shown that
the observed signal evolution is determined by the diffusive motion
of Rb atoms within the porous medium. Our observations are consistent
with the picture of atoms undergoing a one dimensional random walk
along the porous axis. Taking advantage of the well characterized geometry of the porous medium, a simple relation of the diffusion coefficient with the pore diameter and the  atom-wall sticking time was established. Also, at the pores ends, the atom exchange between the gas cell and the porous medium is directly linked, without additional assumptions, to the parameters of the diffusive motion (Eq. \ref{J+}).

The measurement of the diffusion constant gives direct access to the
mean time between steps. This time is essentially a sticking time as
the atoms remain most of the time absorbed to the pore wall. Our results indicate a linear decrease of the sticking time with the
illuminating light intensity for low light intensity. In addition the sticking time
modification appears to be dependent of the illuminating light
frequency. The LIAD yield does not vary monotonically with light
frequency for the three wavelength used. This suggests that the atom
release takes place, at least in part, from rubidium clusters where
surface plasmon resonances contribute to the light absorption
spectrum \cite{Burchianti2006,Burchianti2008}.

\begin{table}
\caption{\label{tab:table} Reported alkali - dielectric surface
sticking times at room temperature}
\begin{ruledtabular}
\begin{tabular}{cccc}

Atom-Surface & Sticking time & Comments \\
\hline

Cs-Pyrex & $1400 \mu$s  & \cite{Stephens1994} \\
Cs-sapphire & $ < 160 \mu$s & \cite{Stephens1994} \\
Na-glass& $130 \mu$s & \cite{Bordo1999}  \\
Rb-alumina & $500 \mu$s &  This work \\
\end{tabular}
\end{ruledtabular}
\end{table}

\section{\label{sec:level14}Acknowledgements}
The authors acknowledge support from ANII (Fondo Clemente Estable), CSIC and ECOS-Sud.\\


\appendix
\section{Calculation of the diffusion coefficient in a cylindrical pore}

For a one dimension random walk in the direction $y$, assuming that the
length and duration of the random steps are uncorrelated, the
diffusion constant is given by \cite{Ott1993}:

\begin{equation}\label{ec:autodifcil}
D = \frac{\left\langle
l_{y}^2\right\rangle} {2\tau}=\frac{\left\langle
v_{y}^2 t^{2}\right\rangle}{2 \tau}
\end{equation}
where $l_{y}$ is the single step displacement in the direction $y$,
$v_{y}$ is the $y$ component of the particle velocity and $t$ the
time-of-flight of a given step. $\tau$  is the mean time interval
between steps.

\begin{figure}[h]
\centering
\includegraphics[width=8.6cm]{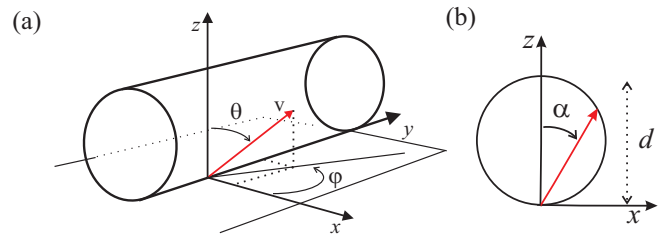}
\caption{(Color online) a) Cylinder and coordinate system considered in the
calculation.  b) Cross-section along the $x,z$
plane.}\label{fig:cilindro}
\end{figure}

We consider particles free flying within the inner surface of
cylinder with diameter $d$. A particle leaving the cylinder wall has
a velocity given by:

\begin{eqnarray}\label{componentes}
\nonumber
    v_{z} &=& v\cos(\theta)\\
    v_{y} &=& v\sin(\theta)\sin(\phi) \\
\nonumber    v_{x} &=& v\sin(\theta)\cos(\phi)
\end{eqnarray}
where $v$ is the velocity modulus. See Fig. \ref{fig:cilindro} for angle definitions.

The time-of-flight is given by:

\begin{equation}\label{ec:tiempProy}
t=\frac{l_{x,z}}{v_{x,z}}
\end{equation}

where $l_{x,z}$ and $v_{x,z}$ are the projections of the particle
displacement and velocity over the $x,z$ plane. We have:

\begin{equation}\label{paso}
l_{x,z}=d \cos(\alpha)=d \frac{v_{z}}{v_{x,z}}
\end{equation}

Using Eqs. \ref{componentes}, \ref{ec:tiempProy} and \ref{paso} we obtain:

\begin{equation}\label{ec:tiempomedio}
t =  \frac{d \cos(\theta)}{v\left(\cos^{2}(\theta)+\sin^{2}(\theta)\cos^{2}(\phi)\right)}\\
\end{equation}

and

\begin{equation}\label{ec:pasomedio}
l_{y}=v_{y}t =
\frac{d\sin(\theta)\cos(\theta)\sin(\phi)}{\left(\cos^{2}(\theta)+\sin^{2}(\theta)\cos^{2}(\phi)\right)}
\end{equation}

The angular (Lambertian) distribution of the atoms leaving the surface is given
by \cite{Goodman1976}:
\begin{equation}\label{distribucionangular}
P(\Omega)d\Omega=\cos(\theta)d\Omega
\end{equation}
where $\Omega$ is the solid angle. The thermal distribution for the
magnitude of the atomic velocity is \cite{Goodman1976}:

\begin{equation}\label{distribucionvelocidad}
P(v) =
\frac{1}{2}\left(\frac{m}{k_{B}T}\right)^{2}v^{3}\exp\left(-\frac{v^{2}}{v_{rms}^{2}}\right)
\end{equation}
with $v_{rms}=\sqrt{\frac{2k_{B}T}{m}}$.\\

Using \ref{ec:pasomedio} and \ref{distribucionangular}, after
integration one gets:

\begin{equation}\label{ec:promediopaso}
\langle l_{y}^{2} \rangle=\frac{2}{3}d^{2}
\end{equation}

In a similar way, from \ref{ec:tiempomedio},
\ref{distribucionangular} and \ref{distribucionvelocidad} we obtain:

\begin{equation}\label{ec:promediopaso}
\tau_{0} = \langle t \rangle=d\sqrt{\frac{2\pi m}{k_{B}T}}
\end{equation}

In our system, the time interval between flights is determined by
the atom sticking time $\tau_{s}$ ( $\tau\simeq \tau_{s} \gg
\tau_{0}$).

\bibliographystyle{apsrev}
\bibliography{refs4}

\end{document}